\documentclass[twocolumn,showpacs,preprintnumbers,nofootinbib,amsmath,amssymb,%
superscriptaddress]{revtex4}

\usepackage{graphicx}
\usepackage{dcolumn}
\usepackage{bm}
\usepackage{epsfig}

\begin{document}

\preprint{DESY~15-131\hspace{14.cm} ISSN~0418-9833}
\preprint{July 2015\hspace{16.9cm}}

\title{\boldmath Stability of the Electroweak Vacuum: Gauge Independence and
Advanced Precision}

\author{A.~V.~Bednyakov}
\affiliation{Joint Institute for Nuclear Research, 141980 Dubna, Russia}
\author{B.~A.~Kniehl}
\affiliation{{II.} Institut f\"ur Theoretische Physik, Universit\"at Hamburg,
Luruper Chaussee 149, 22761 Hamburg, Germany}
\author{A.~F.~Pikelner}
\affiliation{{II.} Institut f\"ur Theoretische Physik, Universit\"at Hamburg,
Luruper Chaussee 149, 22761 Hamburg, Germany}
\author{O.~L.~Veretin}
\affiliation{{II.} Institut f\"ur Theoretische Physik, Universit\"at Hamburg,
Luruper Chaussee 149, 22761 Hamburg, Germany}

\date{\today}

\begin{abstract}
We perform a manifestly gauge-independent analysis of the vacuum stability in
the standard model including two-loop matching, three-loop
renormalization group evolution, and pure QCD corrections through four loops.
All these ingredients are exact, except that light-fermion masses are
neglected.
We in turn apply the criterion of nullifying the Higgs self-coupling and its
beta function in the modified minimal-subtraction scheme and a recently
proposed consistent method for determining the true minimum of the effective
Higgs potential that also avoids gauge dependence.
Exploiting our knowledge of the Higgs-boson mass, we derive an upper bound on
the pole mass of the top quark by requiring that the standard model be stable
all the way up to the Planck mass scale and conservatively estimate the
theoretical uncertainty.
This bound is compatible with the Monte Carlo mass quoted by the Particle Data
Group at the $1.3\sigma$ level.
\end{abstract}

\maketitle

The standard model (SM) of elementary particle physics has been enormously
consolidated by the discovery \cite{Aad:2012tfa} at the CERN Large Hadron
Collider of a new weak neutral resonance that, within the present experimental
uncertainty, shares the spin ($J$), parity ($P$), and charge-conjugation ($C$)
quantum numbers $J^{PC}=0^{++}$ and the coupling strengths with the SM Higgs
boson $H$, in the absence of convincing signals of new physics beyond the SM.
Moreover, its mass of $(125.7\pm0.4)$~GeV \cite{Agashe:2014kda} falls well
inside the $M_H$ range predicted within the SM through global analyses of
electroweak (EW) precision data \cite{Agashe:2014kda}.
Besides completing the SM particle multiplet and confirming the Higgs
mechanism of mass generation via the spontaneous breaking of the EW symmetry
proposed by Englert, Higgs (The Nobel Prize in Physics 2013), and Brout, this
groundbreaking discovery also has fundamental cosmological consequences by
allowing conclusions regarding the fate of the Universe via the analysis of the
vacuum stability \cite{Krasnikov:1978pu}.
In fact, owing to an intriguing conspiracy of the SM particle masses, chances
are that the Higgs potential develops a second minimum, as deep as the one
corresponding to the vacuum with expectation value (VEV)
$v=2^{-1/4}G_F^{-1/2}=246$~GeV in which we live, at a field value of the order
of the Planck mass $M_P=1.22\times10^{19}$~GeV
\cite{Greenwood:2008qp,Bezrukov:2012sa}.
This would imply that the SM be stable all the way up to the energy scale where
the unification with gravity is expected to take place anyways,
which would diminish the necessity for grand unified theories at lower scales.
EW symmetry breaking might thus be determined by Planck-scale physics
\cite{Bezrukov:2012sa}, and the existence of a relationship between $M_P$ and
SM parameters might signify a reduction of fundamental couplings.
Of course, experimental facts that the SM fails to explain, such as the
smallness of the neutrino masses, the strong $CP$ problem, the existence of
dark matter, and the baryon asymmetry in the Universe, would then still call
for an extension.

Obviously, the ultimate answer to the existential question whether our vacuum
is stable or not crucially depends on the quality of the theoretical analysis
as for both conceptual rigor and high precision, and it is the goal of this
Letter to significantly push the state of the art by optimally exploiting
information that has become available just recently.
The technical procedure is as follows.
The set of running coupling constants, including the
SU(2)${}_I$, U(1)${}_Y$, and SU(3)${}_c$ gauge couplings $g(\mu)$,
$g^\prime(\mu)$, and $g_s(\mu)$, respectively, the Higgs self-coupling
$\lambda(\mu)$, and the Yukawa couplings $y_f(\mu)$, of the full SM are evolved
in the renormalization scale $\mu$ from $\mu^\mathrm{thr}=O(v)$ to
$\mu^\mathrm{cri}=O(M_P)$ using the renormalization group (RG)
equations.
The beta functions appearing therein take a simple polynomial form in the
modified minimal-subtraction ($\overline{\mathrm{MS}}$) scheme of dimensional
regularization.
They are fully known through three loops \cite{Mihaila:2012fm} in the
approximation of neglecting the Yukawa couplings of the first- and
second-generation fermions, and the ones of $g_s$ \cite{vanRitbergen:1997va}
and $y_q$ \cite{Chetyrkin:1997dh} also at the four-loop order
$O(\alpha_s^4)$, the latter being given by the quark mass anomalous
dimension.
The initial conditions at $\mu=\mu^\mathrm{thr}$ are evaluated from
the relevant constants of nature, including Sommerfeld's fine-structure
constant $\alpha_\mathrm{Th}$ defined in Thomson scattering---or, alternatively,
Fermi's constant $G_F$---, the strong-coupling constant $\alpha_s^{(5)}(M_Z)$ at
its reference point in QCD with $n_f=5$ active quark flavors, and the physical
particle masses $M_i$ ($i=W,Z,H,f$) defined via the propagator poles, taking
into account threshold corrections \cite{Hempfling:1994ar}, which are fully
known through two loops
\cite{Bezrukov:2012sa,Degrassi:2012ry,Buttazzo:2013uya,Kniehl:2014yia,%
Bednyakov:2014fua,Kniehl:2015nwa}
and, for $g_s$ and $y_q$, also at $O(\alpha_s^3)$
\cite{Chetyrkin:1997sg,Chetyrkin:1999ys} and even at $O(\alpha_s^4)$
\cite{Schroder:2005hy,Marquard:2015qpa}.
Although self-consistency requires that $n$-loop evolution is combined with
$(n-1)$-loop matching, we, nevertheless, include the additional information
\cite{Schroder:2005hy,Marquard:2015qpa} in our default predictions.
There are two approaches to the threshold corrections in the literature that
differ in the definition of the $\overline{\mathrm{MS}}$ VEV $v(\mu)$.
In the first one \cite{Degrassi:2012ry,Buttazzo:2013uya}, $v(\mu)$ is fixed to
be the minimum of the effective Higgs potential $V_\mathrm{eff}(H)$ in the
Landau gauge and is thus gauge dependent \cite{DiLuzio:2014bua}.
A solution to this problem has recently been proposed in
Ref.~\cite{Andreassen:2014gha}.
In the second approach
\cite{Bezrukov:2012sa,Kniehl:2014yia,Bednyakov:2014fua,Kniehl:2015nwa}, the
adjustment of the VEV is only done for the bare theory, yielding
$v^0=\sqrt{-(m_\Phi^0)^2/\lambda^0}$, with $m_\Phi$ being the mass of the
complex scalar doublet $\Phi$, or, equivalently,
\begin{equation}
v^0=\frac{2m_W^0}{e^0}\sqrt{1-\left(\frac{m_W^0}{m_Z^0}\right)^2}
\label{eq:vev}
\end{equation}
in terms of basic parameters of the broken phase \cite{Hempfling:1994ar}.
The linear term in the bare Higgs potential is then quenched and cannot serve
as a tadpole counterterm, so that the tadpole contributions, which carry
gauge dependence, need to be properly included order by order
\cite{Hempfling:1994ar}.
Upon $\overline{\mathrm{MS}}$ renormalization, taking Eq.~(\ref{eq:vev}) with
the superscripts 0 dropped to be exact, $v(\mu)$ and all the basic parameters,
including $\lambda(\mu)$, are manifestly gauge independent to all orders.
Consequently, the twofold vacuum stability condition \cite{Bezrukov:2012sa},
\begin{equation}
\lambda(\mu^\mathrm{cri})=\beta_\lambda(\mu^\mathrm{cri})=0,
\label{eq:crit}
\end{equation}
which fixes a second minimum that is degenerate with the first one,
has gauge-independent solutions for the critical ultrahigh scale
$\mu^\mathrm{cri}$ and one free basic parameter, which we take to be
$M_t^\mathrm{cri}$, the upper bound on the top-quark pole mass $M_t$, which is
much less precisely known than $M_H$ \cite{Agashe:2014kda}.
For comparisons with the literature, we also determine the $M_H$ lower bound
$M_H^\mathrm{cri}$ sloppily using as input the mass parameter
$M_t^\mathrm{MC}$ \cite{Agashe:2014kda} that is extracted from experimental
data using Monte Carlo event generators merely equipped with leading-order (LO)
hard-scattering matrix elements.
The results for $\mu^\mathrm{cri}$ obtained together with $M_t^\mathrm{cri}$
and $M_H^\mathrm{cri}$ are denoted as $\mu_t^\mathrm{cri}$ and
$\mu_H^\mathrm{cri}$, respectively. 
While the criticality condition in Eq.~(\ref{eq:crit}) carries a very simple
physical meaning and is straightforward to solve numerically, it is slightly
scheme dependent.
To assess this scheme dependence, we compare the results for
$\mu_i^\mathrm{cri}$ and $M_i^\mathrm{cri}$ with $i=t,H$ with those obtained
applying the {\it consistent approach} of Ref.~\cite{Andreassen:2014gha}, in
which $V_\mathrm{eff}(H)$ is reorganized in powers of $\hbar$, so that its
expansion coefficients are gauge independent at its extrema
\cite{Nielsen:1975fs}.
Specifically, this amounts to solving
\begin{eqnarray}
\lambda&=&\frac{1}{256\pi^2}\left[
(g^2+g^{\prime2})^2\left(1-3\ln\frac{g^2+g^{\prime2}}{4}\right)
\right.
\nonumber\\
&&{}+\left.
2g^{\prime4}\left(1-3\ln\frac{g^{\prime2}}{4}\right)
-48y_t^4\left(1-\ln\frac{y_t^2}{4}\right)\right],\quad
\end{eqnarray}
which follows from
$dV_\mathrm{eff}^\mathrm{LO}(\tilde{\mu}^\mathrm{cri})/dH=0$, for
the minimum $H=\tilde{\mu}^\mathrm{cri}$ of $V_\mathrm{eff}^\mathrm{LO}(H)$
and requiring that, at next-to-leading order (NLO),
$V_\mathrm{min}^\mathrm{NLO}
=V_\mathrm{eff}^\mathrm{LO}(\tilde{\mu}^\mathrm{cri})
+V_\mathrm{eff}^\mathrm{NLO}(\tilde{\mu}^\mathrm{cri})\ge0$
for $M_t\le \widetilde{M}_t^\mathrm{cri}$ or $M_H\ge \widetilde{M}_H^\mathrm{cri}$,
which is conveniently achieved in the Landau gauge \cite{Buttazzo:2013uya}.

We adopt the input values
$G_F=1.1663787(6)\times10^{-5}$~GeV${}^{-2}$,
$\alpha_s^{(5)}(M_Z)=0.1185(6)$,
$M_W=80.385(15)$~GeV,
$M_Z=91.1876(21)$~GeV,
$M_H=125.7(4)$~GeV,
$M_t^\mathrm{MC}=173.21(87)$~GeV, and
$M_b=4.78(6)$~GeV
from Ref.~\cite{Agashe:2014kda},
evolve $\alpha_s^{(5)}(\mu)$ from $\mu=M_Z$ to the matching scale
$\mu^\mathrm{thr}=\xi M_t^\mathrm{MC}$ in the $n_f=5$ effective theory using
coupled QCD${}\times{}$QED beta functions through four loops in QCD
\cite{vanRitbergen:1997va} and three loops in QED \cite{Erler:1998sy}, and
evaluate there the $\overline{\mathrm{MS}}$ couplings of the full SM from
\begin{eqnarray}
g^2(\mu)&=&2^{5/2}G_FM_W^2[1+\delta_W(\mu)],
\nonumber\\
g^2(\mu)+g^{\prime2}(\mu)&=&2^{5/2}G_FM_Z^2[1+\delta_Z(\mu)],
\nonumber\\
\lambda(\mu)&=&2^{-1/2}G_FM_H^2[1+\delta_H(\mu)],
\nonumber\\
y_f(\mu)&=&2^{3/4}G_F^{1/2}M_f[1+\delta_f(\mu)],
\nonumber\\
g_s^2(\mu)&=&4\pi\alpha_s^{(5)}(\mu)[1+\delta_{\alpha_s}(\mu)],
\label{eq:thresh}
\end{eqnarray}
including the appropriate terms of
$O(\alpha^n)$ with $n=1,2$ \cite{Kniehl:2014yia,Kniehl:2015nwa},
$O(\alpha\alpha_s)$
\cite{Bezrukov:2012sa,Kniehl:2014yia,Bednyakov:2014fua,Kniehl:2015nwa}, and
$O(\alpha_s^n)$ with $n=1,2,3,4$
\cite{Chetyrkin:1997sg,Chetyrkin:1999ys,Schroder:2005hy,Marquard:2015qpa}.
The threshold corrections $\delta_i(\mu)$ in Eq.~(\ref{eq:thresh}) are
expressed in terms of the $\overline{\mathrm{MS}}$ couplings $\alpha(\mu)$ and
$\alpha_s(\mu)$, and the pole masses $M_i$.
To avoid the theoretical uncertainty due to the hadronic contributions to the
relationship between $\alpha(\mu)$ and $\alpha_\mathrm{Th}$
\cite{Agashe:2014kda}, we replace the latter by $G_F$ in the set of basic
parameters by extracting $\alpha(\mu)$ self-consistently from the exact
relationship $1/[4\pi\alpha(\mu)]=1/g^2(\mu)+1/g^{\prime2}(\mu)$
\cite{Kniehl:2015nwa}.
We stress that the $\overline{\mathrm{MS}}$ couplings in Eq.~(\ref{eq:thresh})
are manifestly gauge independent and, thanks to partial tadpole cancellations,
also finite in the limit $M_H\to0$ \cite{Kniehl:2014yia,Kniehl:2015nwa}.
The pole masses $M_t$ and $M_b$ are subject to renormalon ambiguities of
$O(\Lambda_\mathrm{QCD})$, which, for $M_t$, are still small against
the experimental error \cite{Agashe:2014kda} and, for $M_b$, are
inconsequential because of the smallness of $y_b(\mu)$.
The use of $\overline{\mathrm{MS}}$ masses $m_q(\mu)$ would avoid renormalon
ambiguities at the expense of introducing unscreened tadpole contributions to
restore gauge independence \cite{Hempfling:1994ar}, which coincidentally
reduce the scheme dependence of $m_t(\mu)$ \cite{Jegerlehner:2012kn}, but spoil
the perturbative expansion for $m_b(\mu)$ \cite{Kniehl:2014yia}.
For completeness, we also study the $\overline{\mathrm{MS}}$ mass parameter of
the Higgs potential,
$m^2(\mu)=-2m_\Phi^2(\mu)=2v^2(\mu)\lambda(\mu)$, using
\begin{equation}
v^2(\mu)=2^{-1/2}G_F^{-1}[1+\Delta\bar{r}(\mu)],
\end{equation}
where $\Delta\bar{r}(\mu)$ to $O(\alpha^n)$ with $n=1,2$ and
$O(\alpha\alpha_s)$ may be found in Ref.~\cite{Kniehl:2015nwa}.
$\Delta\bar{r}(\mu)$ is gauge independent, but diverges for $M_H\to0$ due to
unscreened tadpole contributions.
We estimate the theoretical uncertainties in the $\overline{\mathrm{MS}}$
parameters for $\xi=1$ due to unknown higher-order corrections by considering
both scale variations and truncation errors.
In the first case, we in turn put $\xi=1/2$ and 2 in Eq.~(\ref{eq:thresh}),
return to $\xi=1$ using the RG equations, and select the larger one of the two
deviations thus generated.
In the second case, we find the full set of $\overline{\mathrm{MS}}$
parameters for $\xi=1$, including besides those in Eq.~(\ref{eq:thresh}) also
$m_i(\mu)$ with $i=W,Z,H,f$ and $v(\mu)$, by self-consistently solving the
system of equations that express $G_F$
and
$M_i$ entirely in terms of these parameters, so that unscreened tadpole
contributions have to cancel numerically.
\begin{table*}[t]
 \centering
\begin{tabular}{ccccccccc}
\hline
\hline
$x$ & $x_0$ & $\Delta x_{\alpha_s}$ & $\Delta x_{M_W}$ & $\Delta x_{M_H}$ &
$\Delta x_{M_t}$ & $\beta_x$ & $\delta x_\mu$ & $\delta x_\mathrm{tru}$ \\
\hline
$g$ & $0.35838$ & $-3.8\times 10^{-6}$ & $-2.3\times 10^{-4}$ & 
$-2.5\times 10^{-6}$ & $+7.1\times 10^{-5}$ & $+2.1\times 10^{-3}$ &
$8.5\times 10^{-5}$ &  $6.4\times 10^{-4}$ \\
$g^\prime$ & $0.64812$ & $+8.5\times 10^{-7}$ & $+1.2\times 10^{-4}$ &
$-6.6\times 10^{-7}$ & $-9.8\times 10^{-6}$ &  $-5.2\times 10^{-3}$ &
$5.8\times 10^{-5}$ &  $1.0\times 10^{-3}$ \\
$g_s$ & $1.16540$ & $+2.7\times 10^{-3}$ & $+8.9\times 10^{-8}$ &
$+7.8\times 10^{-8}$ & $-4.0\times 10^{-5}$ & $-7.2\times 10^{-2}$ &
$5.6\times 10^{-5}$ & $\cdots$ \\
$y_t$ & $0.93517$ & $-3.6\times 10^{-4}$ & $-1.3\times 10^{-7}$ &
$-8.6\times 10^{-6}$ & $+5.1\times 10^{-3}$ & $-5.2\times 10^{-2}$ &
$8.0\times 10^{-4}$ & $1.2\times 10^{-3}$ \\
$y_b$ & $0.01706$ & $-5.7\times 10^{-5}$ & $-5.1\times 10^{-10}$ &
$+1.3\times 10^{-7}$ & $-2.4\times 10^{-7}$ & $-9.2\times 10^{-4}$ &
$2.5\times 10^{-4}$ & $1.1\times 10^{-3}$ \\
$\lambda$ & $0.12714$ & $-6.2\times 10^{-6}$ & $-4.2\times 10^{-7}$ &
$+8.2\times 10^{-4}$ & $+6.4\times 10^{-5}$ & $-2.0\times 10^{-2}$ &
$5.8\times 10^{-4}$ & $5.5\times 10^{-4}$ \\
$m$ & $131.86$ & $-2.6\times 10^{-3}$ & $-4.4\times 10^{-4}$ &
$+3.8\times 10^{-1}$ & $+1.2\times 10^{-1}$ & $+2.6$ &
$7.3\times 10^{-1}$ & $4.1\times 10^{-2}$ \\
\hline
\hline
\end{tabular}
\caption{\label{tab:x}
Coefficients in Eq.~(\ref{eq:x}).
The entries in the last row are given in units of GeV.}
\end{table*}
We cast our results for $x=g,g^\prime,g_s,y_t,y_b,\lambda,m$ in the form
\begin{eqnarray}
x(\mu)&=&x_0
+\Delta x_{\alpha_s}\frac{\alpha_s^{(5)}(M_Z)-\alpha_s^{(5),\mathrm{exp}}(M_Z)}
{\Delta\alpha_s^{(5),\mathrm{exp}}(M_Z)}
\nonumber\\
&&{}+\Delta x_{M_W}\frac{M_W-M_W^\mathrm{exp}}{\Delta M_W^\mathrm{exp}}
+\Delta x_{M_H}\frac{M_H-M_H^\mathrm{exp}}{\Delta M_H^\mathrm{exp}}
\nonumber\\
&&{}+\Delta x_{M_t}\frac{M_t-M_t^\mathrm{exp}}{\Delta M_t^\mathrm{exp}}
+\beta_x\frac{\mu-\mu^\mathrm{thr}}{\mu^\mathrm{thr}}
\pm\delta x_\mu
\nonumber\\
&&{}\pm\delta x_\mathrm{tru},
\label{eq:x}
\end{eqnarray}
allowing for linear extrapolations in the least precisely known input
parameters quoted above \cite{Agashe:2014kda} and $\mu^\mathrm{thr}$, which we
disentangle from $M_t^\mathrm{exp}=M_t^\mathrm{MC}$, and list the coefficients in
Table~\ref{tab:x}.

\begin{table*}[t]
 \centering
\begin{tabular}{ccccccccccc}
\hline
\hline
$X$ & $X_0$ & $\Delta X_{\alpha_s}$ & $\Delta X_M$ & $\delta X_\mathrm{par}$ &
$\delta X_\mu^+$ & $\delta X_\mu^-$ & $\delta X_\mathrm{tru}$ &
$\delta_i^{O(\alpha^2)}$ &
$\delta_{\alpha_s}^{O(\alpha\alpha_s,\alpha_s^4)}$ &
$\delta_q^{O(\alpha_s^4)}$ \\
\hline
$M_t^\mathrm{cri}$ & $171.44$ & $0.23$ & $0.20$ & $0.001$ &
$-0.36$ & $0.17$ & $-0.02$ &
$171.55_{+1.04}^{-0.47}$ &
$171.43_{+0.17}^{-0.36}$ &
$171.24_{+0.19}^{-0.38}$ \\
$\log_{10}\mu_t^\mathrm{cri}$ & $17.752$ & $-0.051$ & $0.083$ & $0.007$ & 
$0.007$ & $-0.006$ & $-0.002$ &
$17.783_{-0.008}^{+0.062}$ &
$17.754_{-0.006}^{+0.007}$ &
$17.751_{-0.007}^{+0.007}$ \\
$M_H^\mathrm{cri}$ & $129.30$ & $-0.49$ & $1.79$ & $0.002$ &
$0.72$ & $-0.33$ & $0.04$ &
$129.06_{-2.14}^{+0.95}$ &
$129.32_{-0.33}^{+0.73}$ &
$129.72_{-0.38}^{+0.76}$ \\
$\log_{10}\mu_H^\mathrm{cri}$ & $18.512$ & $-0.158$ & $0.381$ & $0.008$ & 
$0.173$ & $-0.082$ & $0.008$ &
$18.495_{-0.531}^{+0.226}$ &
$18.518_{-0.082}^{+0.174}$ &
$18.602_{-0.094}^{+0.184}$ \\
$\widetilde{M}_t^\mathrm{cri}$ & $171.64$ & $0.23$ & $0.20$ & $0.001$ &
$-0.36$ & $0.17$ & $-0.02$ &
$171.74_{+1.04}^{-0.46}$ &
$171.63_{+0.17}^{-0.36}$ &
$171.43_{+0.19}^{-0.37}$ \\
$\log_{10}\tilde{\mu}_t^\mathrm{cri}$ & $21.442$ & $-0.059$ & $0.094$ & $0.005$ & 
$-0.083$ & $0.022$ & $0.002$ &
$21.485_{+0.343}^{-0.085}$ &
$21.445_{+0.022}^{-0.083}$ &
$21.441_{+0.014}^{-0.072}$ \\
$\widetilde{M}_H^\mathrm{cri}$ & $128.90$ & $-0.49$ & $1.79$ & $0.003$ &
$0.73$ & $-0.34$ & $0.04$ &
$128.67_{-2.15}^{+0.95}$ &
$128.92_{-0.34}^{+0.73}$ &
$129.32_{-0.38}^{+0.76}$ \\
$\log_{10}\tilde{\mu}_H^\mathrm{cri}$ & $22.209$ & $-0.181$ & $0.436$ & $0.007$ & 
$0.092$ & $-0.062$ & $0.013$ &
$22.201_{-0.171}^{+0.146}$ &
$22.217_{-0.062}^{+0.094}$ &
$22.312_{-0.082}^{+0.113}$ \\
\hline
\hline
\end{tabular}
\caption{\label{tab:X}
Coefficients in Eq.~(\ref{eq:X}) and central values with scale dependencies
obtained upon switching off the $O(\alpha^2)$ terms in
$\delta_i(\mu)$ with $i=W,Z,H,q$, the $O(\alpha\alpha_s)$ and
$O(\alpha_s^4)$ terms in $\delta_{\alpha_s}(\mu)$, and the
$O(\alpha_s^4)$ terms in $\delta_q(\mu)$ one at a time.
The unit of mass is taken to be GeV.}
\end{table*}
We now in turn apply criterion~(\ref{eq:crit}) and the approach of
Ref.~\cite{Andreassen:2014gha} and write the resulting critical masses
and associated scales 
$X=M_i^\mathrm{cri},\mu_i^\mathrm{cri},\widetilde{M}_i^\mathrm{cri},
\tilde{\mu}_i^\mathrm{cri}$
with $i=t,H$ in the form
\begin{eqnarray}
X&=&X_0
+\Delta X_{\alpha_s}\frac{\alpha_s^{(5)}(M_Z)-\alpha_s^{(5),\mathrm{exp}}(M_Z)}
{\Delta\alpha_s^{(5),\mathrm{exp}}(M_Z)}
\nonumber\\
&&{}+\Delta X_M\frac{M-M^\mathrm{exp}}{\Delta M^\mathrm{exp}}
\pm\delta X_\mathrm{par}
+\delta X_\mu^\pm
\pm\delta X_\mathrm{tru},
\quad\label{eq:X}
\end{eqnarray}
where $M=M_H$ ($M_t$) if $i=t$ ($H$),
$\Delta X_{\alpha_s}$ and $\Delta X_M$ are the $1\sigma$ errors due to
$\alpha_s^{(5)}(M_Z)$ and $M$, respectively,
$\delta X_\mathrm{par}$ are the residual parametric errors combined in
quadrature,
$\delta X_\mu^\pm$ are the shifts due to the choices $\xi=2^{\pm1}$, and
$\delta X_\mathrm{tru}$ are the truncation errors induced by those in
Table~\ref{tab:x}.
The coefficients in Eq.~(\ref{eq:X}) are collected in Table~\ref{tab:X}.
$\widetilde{M}_t^\mathrm{cri}$ is $0.20$~GeV larger than $M_t^\mathrm{cri}$, and
$\widetilde{M}_H^\mathrm{cri}$ is $0.40$~GeV smaller than $M_H^\mathrm{cri}$.
These shifts reflect the scheme dependence.
$\mu_t^\mathrm{cri}$ and $\mu_H^\mathrm{cri}$ fall slightly short of $M_P$,
for which $\log_{10}M_P=19.086$, where the SM definitely ceases to be valid,
while $\tilde{\mu}_t^\mathrm{cri}$ and $\tilde{\mu}_H^\mathrm{cri}$ lie
appreciably beyond $M_P$, which is an inherent problem of
Ref.~\cite{Andreassen:2014gha} and was cured there by the {\it ad hoc}
introduction of some new dimension-six operator.
In the remainder of this Letter, we concentrate on the approach based on
Eq.~(\ref{eq:crit}) \cite{Bezrukov:2012sa}.

To assess the significance of the higher-order corrections that were not yet
included in Ref.~\cite{Bezrukov:2012sa}, namely the full
$O(\alpha^2)$ terms in $\delta_i(\mu)$ with $i=W,Z,H,q$
\cite{Kniehl:2014yia,Kniehl:2015nwa},
the $O(\alpha\alpha_s)$ term in $\delta_{\alpha_s}(\mu)$
\cite{Bednyakov:2014fua}, and
the $O(\alpha_s^4)$ terms in $\delta_{\alpha_s}(\mu)$
\cite{Schroder:2005hy} and $\delta_q(\mu)$ \cite{Marquard:2015qpa},
we switch them off one at a time.
The resulting central values and scale dependencies of the critical parameters
are also contained in Table~\ref{tab:X}.
The $O(\alpha^2)$ terms in $\delta_i(\mu)$
\cite{Kniehl:2014yia,Kniehl:2015nwa} shift $M_t^\mathrm{cri}$ and
$M_H^\mathrm{cri}$ by $-0.11$ and $+0.24$~GeV, respectively, and reduce
their scale uncertainties by almost a factor of 3.
On the other hand, the $O(\alpha_s^4)$ terms in $\delta_q(\mu)$
\cite{Marquard:2015qpa} produce larger and opposite shifts in
$M_t^\mathrm{cri}$ and $M_H^\mathrm{cri}$, namely $+0.20$ and $-0.42$~GeV,
respectively, but merely reduce their scale uncertainties by less than 10\%.
The $O(\alpha\alpha_s)$ \cite{Bezrukov:2012sa} and
$O(\alpha_s^4)$ \cite{Schroder:2005hy} terms in
$\delta_{\alpha_s}(\mu)$ are much less significant.
All these observations approximately carry over to
$\widetilde{M}_t^\mathrm{cri}$ and $\widetilde{M}_H^\mathrm{cri}$.

Apart from the issue of gauge dependence, our analysis differs from that of
Refs.~\cite{Degrassi:2012ry,Buttazzo:2013uya} in the following respects.
In Refs.~\cite{Degrassi:2012ry,Buttazzo:2013uya}, the 
$O(\alpha\alpha_s)$ term in $\delta_{\alpha_s}(\mu)$
\cite{Bednyakov:2014fua} and the $O(\alpha_s^4)$ terms in
$\delta_{\alpha_s}(\mu)$ \cite{Schroder:2005hy} and $\delta_q(\mu)$
\cite{Marquard:2015qpa} were not included;
$\mu^\mathrm{thr}$ was affected by the $M_t^\mathrm{MC}$ variation, which
explains the sign difference in the corresponding shift in $M_H^\mathrm{cri}$;
and the scale uncertainties were found to be approximately half as large as
here for reasons unknown to us.

\begin{figure}[th]
\centering
\includegraphics[width=0.48\textwidth]{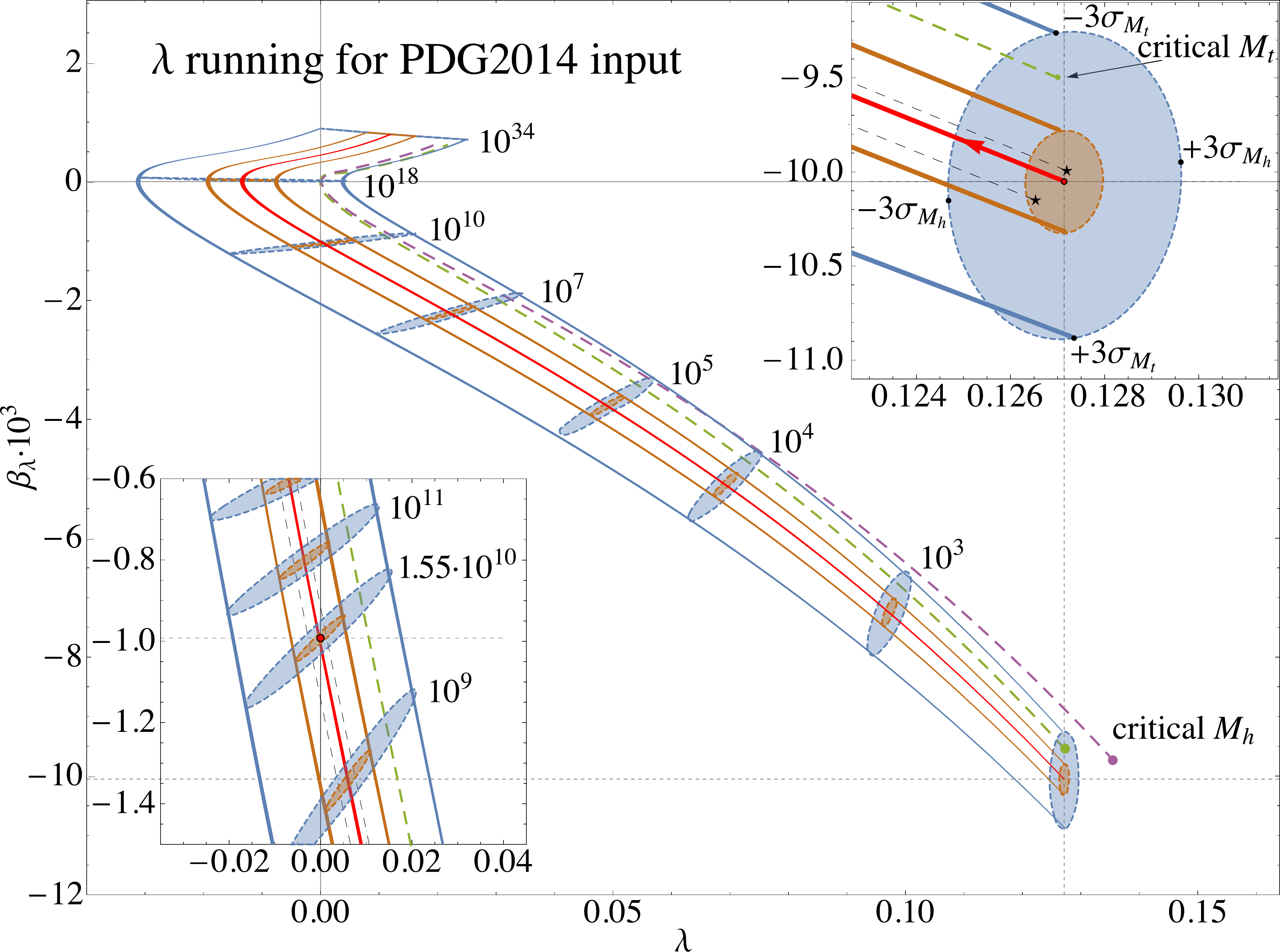}
\caption{\label{fig:blam_lam_evo}%
RG evolution of $\lambda(\mu)$ from $\mu^\mathrm{thr}$ to $\mu^\mathrm{cri}$
and beyond in the $(\lambda,\beta_\lambda)$ plane for default input values
and matching scale (red solid line),
effects of $1\sigma$ (brown solid lines) and $3\sigma$ (blue solid lines)
variation in $M_t^\mathrm{MC}$,
theoretical uncertainty due to the variation of $\xi$ from $1/2$ to 2 
(upper and lower black dashed lines with asterisks in the insets),
and results for $M_t^\mathrm{cri}$ (green dashed line) and
$M_H^\mathrm{cri}$ (purple dashed line).
The $1\sigma$ (brown ellipses) and $3\sigma$ (blue ellipses) contours due to
the errors in $M_t^\mathrm{MC}$ and $M_H$ are indicated for selected values of
$\mu$.
The insets in the upper right and lower left corners refer to
$\mu=M_t^\mathrm{MC}$ and $\mu=1.55\times10^{10}$~GeV, respectively.}
\end{figure}
In Fig.~\ref{fig:blam_lam_evo}, the RG evolution flow from $\mu^\mathrm{thr}$
to $\mu^\mathrm{cri}$ and beyond is shown in the $(\lambda,\beta_\lambda)$
plane.
The propagation with $\mu$ of the $1\sigma$ and $3\sigma$ confidence ellipses
with respect to $M_t^\mathrm{MC}$ and $M_H$ tells us that the second condition
in Eq.~(\ref{eq:crit}) is almost automatic, the ellipses for $\mu=10^{18}$~GeV
being approximately degenerated to horizontal lines.
For default input values, $\lambda(\mu)$ crosses zero at
$\mu=1.55\times10^{10}$~GeV.
The contour of $M_t^\mathrm{cri}$ approximately coincides with the right
envelope of the $2\sigma$ ellipses, while the one of $M_H^\mathrm{cri}$, which
relies on $M_t^\mathrm{MC}$, is driven outside the $3\sigma$ band as $\mu$ runs
from $\mu_H^\mathrm{cri}$ to $\mu^\mathrm{thr}$.

\begin{figure}[th]
\centering
\includegraphics[width=0.52\textwidth]{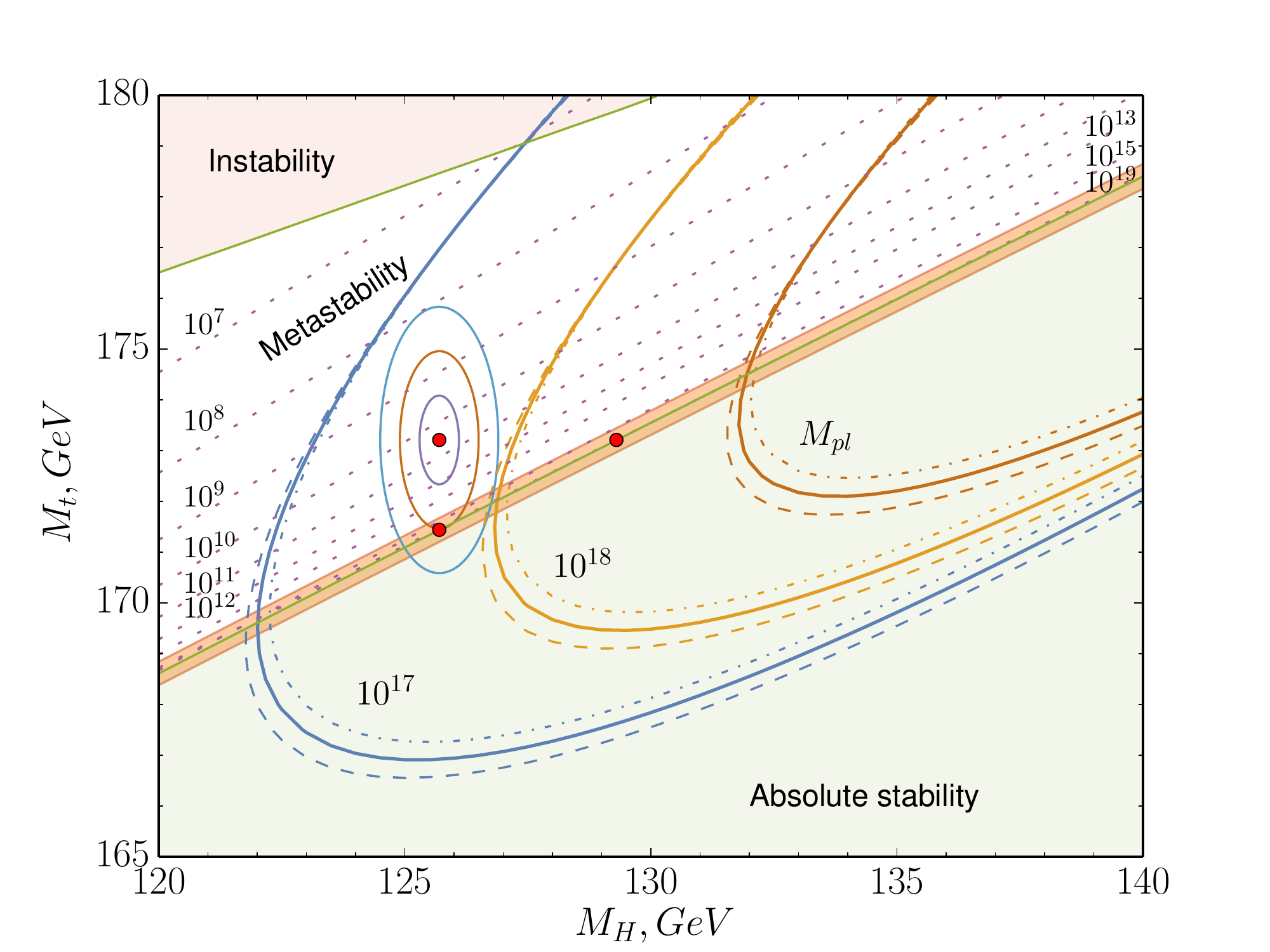}
\caption{\label{fig:mt_mh_phase}%
Phase diagram of vacuum stability (light-green shaded area), metastability,
and instability (pink shaded area) in the $(M_H,M_t)$ plane,
contours of $\lambda(\mu^0)=0$ for selected values of $\mu^0$ (purple dotted
lines),
contours of $\beta_\lambda(\mu^0)=0$ for selected values of $\mu^0$ (solid
parabolalike lines) with uncertainties due to $1\sigma$ error in
$\alpha_s^{(5)}(M_Z)$ (dashed and dot-dashed lines),
critical line of Eq.~(\ref{eq:crit}) (solid green line) with uncertainty due to
$1\sigma$ error in $\alpha_s^{(5)}(M_Z)$ (orange shaded band), and
critical points with $M_t^\mathrm{cri}$ (lower red bullet) and $M_H^\mathrm{cri}$
(right red bullet).
The present world average of $(M_t^\mathrm{MC},M_H)$ (upper left red bullet)
and its $1\sigma$ (purple ellipse), $2\sigma$ (brown ellipse), and $3\sigma$
(blue ellipse) contours are marked for reference.}
\end{figure}

Our upgraded and updated version of the familiar phase diagram
\cite{Degrassi:2012ry,Buttazzo:2013uya,Andreassen:2014gha,Alekhin:2012py} is
presented in Fig.~\ref{fig:mt_mh_phase}.
Besides the boundary of the stable phase defined by Eq.~(\ref{eq:crit}), on
which the critical points with $M_t^\mathrm{cri}$ and $M_H^\mathrm{cri}$ are
located, we also show contours of $\lambda(\mu^0)=0$ and
$\beta_\lambda(\mu^0)=0$.
The demarcation line between the metastable phase and the instable one, in
which the lifetime of our vacuum is shorter than the age of the Universe, is
evaluated as in Ref.~\cite{Andreassen:2014gha} and represents the only
gauge-dependent detail in Fig.~\ref{fig:mt_mh_phase}.
The customary confidence ellipses with respect to $M_t^\mathrm{MC}$ and
$M_H$, which are included Fig.~\ref{fig:mt_mh_phase} for reference, have to be
taken with caution because they misleadingly suggest that the
tree-level mass parameter $M_t^\mathrm{MC}$ and its error \cite{Agashe:2014kda}
identically carry over to $M_t$, which is actually the real part of the
complex pole position upon mass renormalization in the on-shell scheme
\cite{Kniehl:2008cj}.
In view of the resonance property, a shift of order $\Gamma_t=2.00$~GeV
\cite{Agashe:2014kda} would be plausible, which should serve as a useful
error estimate for the time being.

In conclusion, we performed a high-precision analysis of the vacuum stability
in the SM incorporating full two-loop threshold corrections
\cite{Bezrukov:2012sa,Kniehl:2014yia,Bednyakov:2014fua,Kniehl:2015nwa},
three-loop beta functions \cite{Mihaila:2012fm}, and $O(\alpha_s^4)$
corrections to the matching and running of $g_s$
\cite{vanRitbergen:1997va,Schroder:2005hy} and $y_q$
\cite{Chetyrkin:1997dh,Marquard:2015qpa}, and adopting two gauge-independent
approaches, one based on the criticality criterion~(\ref{eq:crit}) for
$\lambda(\mu)$ \cite{Bezrukov:2012sa} and one on a reorganization of
$V_\mathrm{eff}(H)$ so that its minimum is gauge independent order by order
\cite{Andreassen:2014gha}.
For the $M_t$ upper bound we thus obtained
$M_t^\mathrm{cri}=(171.44\pm0.30{+0.17\atop-0.36})$~GeV and
$\widetilde{M}_t^\mathrm{cri}=(171.64\pm0.30{+0.17\atop-0.36})$~GeV,
respectively, where the first errors are experimental, due the $1\sigma$
variations in the input parameters \cite{Agashe:2014kda}, and the second ones
are theoretical, due to the scale and truncation uncertainties.
In want of more specific information, we assume the individual error sources
to be independent and combine them quadratically to be on the conservative
side.
The $0.20$~GeV difference between the central values of $M_t^\mathrm{cri}$ and
$\widetilde{M}_t^\mathrm{cri}$ indicates the scheme dependence, which arguably
comes as a third independent source of theoretical uncertainty.
As our final result, we hence quote the combined value
$\widehat{M}_t^\mathrm{cri}=(171.54\pm0.30{+0.26\atop-0.41})$~GeV, which
is compatible
with $M_t^\mathrm{MC}=(173.21\pm0.87)$~GeV at the $1.3\sigma$ level.
In view of this and the present lack of knowledge of the precise relationship
between and $M_t^\mathrm{MC}$ and $M_t$ mentioned above, the familiar notion
\cite{Degrassi:2012ry,Buttazzo:2013uya} that our vacuum is metastable is
likely to be premature \cite{Alekhin:2012py}.

We thank W.~Frost and M.~D.~Schwartz for useful correspondence regarding
Ref.~\cite{Andreassen:2014gha} and M.~Yu.~Kalmykov for critical comments.
This work was supported in part by DFG Grant No.\ SFB~676, MES of Russia Grant
No.\ MK--1001.2014.2, the Heisenberg-Landau Programme, and the Dynasty
Foundation.


\begin{thebibliography}{99}

\bibitem{Aad:2012tfa}
  G.~Aad {\it et al.}\ (ATLAS Collaboration),
  Phys.\ Lett.\ B {\bf 716}, 1 (2012);
  S.~Chatrchyan {\it et al.}\ (CMS Collaboration),
{\it ibid.}\ {\bf 716}, 30 (2012).

\bibitem{Agashe:2014kda} 
  K.~A.~Olive {\it et al.}\  (Particle Data Group),
  Chin.\ Phys.\ C {\bf 38}, 090001 (2014).

\bibitem{Krasnikov:1978pu}
  N.~V.~Krasnikov,
  Yad.\ Fiz.\ {\bf 28}, 549  (1978)
  [Sov.\ J. Nucl.\ Phys.\ {\bf 28}, 279 (1978)];
  H.~D.~Politzer and S.~Wolfram,
  Phys.\ Lett.\ B {\bf 82}, 242 (1979);
  {\bf 83}, 421(E) (1979);
  P.~Q.~Hung,
  Phys.\ Rev.\ Lett.\ {\bf 42}, 873 (1979).

\bibitem{Greenwood:2008qp} 
  E.~Greenwood, E.~Halstead, R.~Poltis, and D.~Stojkovic,
  Phys.\ Rev.\ D {\bf 79}, 103003 (2009).

\bibitem{Bezrukov:2012sa}
  F.~Bezrukov, M.~Yu.~Kalmykov, B.~A.~Kniehl, and M.~Shaposhnikov,
  J. High Energy Phys.\ 10 (2012) 140.

\bibitem{Mihaila:2012fm} 
  L.~N.~Mihaila, J.~Salomon, and M.~Steinhauser,
  Phys.\ Rev.\ Lett.\  {\bf 108}, 151602 (2012);
  Phys.\ Rev.\ D {\bf 86}, 096008 (2012);
  K.~G.~Chetyrkin and M.~F.~Zoller,
  J. High Energy Phys.\ 06 (2012) 033;
  04 (2013) 091;
  09 (2013) 155(E);
  A.~V.~Bednyakov, A.~F.~Pikelner, and V.~N.~Velizhanin,
   {\it ibid.}\ 01 (2013) 017;
  Phys.\ Lett.\ B {\bf 722}, 336 (2013);
  Nucl.\ Phys.\ {\bf B875}, 552 (2013).

\bibitem{vanRitbergen:1997va} 
  T.~van Ritbergen, J.~A.~M.~Vermaseren, and S.~A.~Larin,
  Phys.\ Lett.\ B {\bf 400}, 379 (1997);
  M.~Czakon,
  Nucl.\ Phys.\ {\bf B710}, 485 (2005).

\bibitem{Chetyrkin:1997dh} 
  K.~G.~Chetyrkin,
  Phys.\ Lett.\ B {\bf 404}, 161 (1997);
  Nucl.\ Phys.\ {\bf B710}, 499 (2005);
  J.~A.~M.~Vermaseren, S.~A.~Larin, and T.~van Ritbergen,
  Phys.\ Lett.\ B {\bf 405}, 327 (1997).

\bibitem{Hempfling:1994ar} 
  R.~Hempfling and B.~A.~Kniehl,
  Phys.\ Rev.\ D {\bf 51}, 1386 (1995).

\bibitem{Degrassi:2012ry}
  G.~Degrassi, S.~Di Vita, J.~Elias-Mir\'o, J.~R.~Espinosa, G.~F.~Giudice,
  G.~Isidori, and A.~Strumia,
  J. High Energy Phys.\ 08 (2012) 098.

\bibitem{Buttazzo:2013uya}
  D.~Buttazzo, G.~Degrassi, P.~P.~Giardino, G.~F.~Giudice, F.~Sala, A.~Salvio,
  and A.~Strumia,
  J. High Energy Phys.\ 
  12 (2013) 089.

\bibitem{Kniehl:2014yia} 
  B.~A.~Kniehl and O.~L.~Veretin,
  Nucl.\ Phys.\ {\bf B885}, 459 (2014);
  {\bf B894}, 56(E) (2015).

\bibitem{Bednyakov:2014fua} 
  A.~V.~Bednyakov,
  Phys.\ Lett.\ B {\bf 741}, 262 (2015).

\bibitem{Kniehl:2015nwa} 
  B.~A.~Kniehl, A.~F.~Pikelner, and O.~L.~Veretin,
  Nucl.\ Phys.\ {\bf B896}, 19 (2015).

\bibitem{Chetyrkin:1997sg} 
  K.~G.~Chetyrkin, B.~A.~Kniehl, and M.~Steinhauser,
  Phys.\ Rev.\ Lett.\  {\bf 79}, 2184 (1997);
  Nucl.\ Phys.\ {\bf B510}, 61 (1998).

\bibitem{Chetyrkin:1999ys} 
  K.~G.~Chetyrkin and M.~Steinhauser,
  Phys.\ Rev.\ Lett.\  {\bf 83}, 4001 (1999);
  Nucl.\ Phys.\ {\bf B573}, 617 (2000);
  K.~Melnikov and T.~van~Ritbergen,
  Phys.\ Lett.\ B {\bf 482}, 99 (2000).

\bibitem{Schroder:2005hy} 
  Y.~Schr\"oder and M.~Steinhauser,
  J. High Energy Phys.\ 01 (2006) 051;
  K.~G.~Chetyrkin, J.~H.~K\"uhn, and C.~Sturm,
  Nucl.\ Phys.\ {\bf B744}, 121 (2006);
  B.~A.~Kniehl, A.~V.~Kotikov, A.~I.~Onishchenko, and O.~L.~Veretin,
  Phys.\ Rev.\ Lett.\  {\bf 97}, 042001 (2006).

\bibitem{Marquard:2015qpa} 
  P.~Marquard, A.~V.~Smirnov, V.~A.~Smirnov, and M.~Steinhauser,
  Phys.\ Rev.\ Lett.\  {\bf 114}, 142002 (2015).

\bibitem{DiLuzio:2014bua} 
  L.~Di Luzio and L.~Mihaila,
  J. High Energy Phys.\ 06 (2014) 079.

\bibitem{Andreassen:2014gha} 
  A.~Andreassen, W.~Frost, and M.~D.~Schwartz,
  Phys.\ Rev.\ Lett.\  {\bf 113}, 241801 (2014);
  Phys.\ Rev.\ D {\bf 91}, 016009 (2015).

\bibitem{Nielsen:1975fs} 
  N.~K.~Nielsen,
  Nucl.\ Phys.\ {\bf B101}, 173 (1975);
  R.~Fukuda and T.~Kugo,
  Phys.\ Rev.\ D {\bf 13}, 3469 (1976).

\bibitem{Erler:1998sy} 
  J.~Erler,
  Phys.\ Rev.\ D {\bf 59}, 054008 (1999).

\bibitem{Jegerlehner:2012kn} 
  F.~Jegerlehner, M.~Yu.~Kalmykov, and B.~A.~Kniehl,
  Phys.\ Lett.\ B {\bf 722}, 123 (2013).

\bibitem{Alekhin:2012py} 
  S.~Alekhin, A.~Djouadi, and S.~Moch,
  Phys.\ Lett.\ B {\bf 716}, 214 (2012).

\bibitem{Kniehl:2008cj} 
  B.~A.~Kniehl and A.~Sirlin,
  Phys.\ Rev.\ D {\bf 77}, 116012 (2008);
{\bf 85}, 036007 (2012);
  B.~A.~Kniehl,
  Phys.\ Rev.\ Lett.\  {\bf 112}, 071603 (2014);
  Phys.\ Rev.\ D {\bf 89}, 096005 (2014);
{\bf 89}, 116010 (2014).

\end{thebibliography}
\end{document}